\documentclass[twocolumn,showpacs,amsmath,amssymb,eqsecnum]{revtex4}

\usepackage{graphicx}

\begin{document}

\title{Canonical theory of spherically
symmetric spacetimes with cross-streaming null dusts}

\author{Ji\v{r}\'{\i} Bi\v{c}\'{a}k}
\affiliation{%
Institute of Theoretical Physics, Charles University,\\
V Hole\v{s}ovi\v{c}k\'{a}ch 2, 182 00 Prague 8,\\
Czech Republic%
}
\author{Petr H\'{a}j\'{\i}\v{c}ek}%
\affiliation{%
Institute of Theoretical Physics, University of Berne,\\
Berne, Switzerland\\
}

\date{August 6, 2003} 

\begin{abstract}
  The Hamiltonian dynamics of two-component spherically
  symmetric null dust is studied with regard to the quantum theory of
  gravitational collapse. The components---the ingoing and outgoing
  dusts---are assumed to interact only through gravitation.  Different kinds
  of singularities, naked or ``clothed'', that can form during collapse
  processes are described. The general canonical formulation of the
  one-component null-dust dynamics by Bi\v{c}\'{a}k and Kucha\v{r} is
  restricted to the spherically symmetric case and used to construct an action
  for the two components. The transformation from a metric variable to the
  quasilocal mass is shown to simplify the mathematics. The action is reduced
  by a choice of gauge and the corresponding true Hamiltonian is written down.
  Asymptotic coordinates and energy densities of dust shells are shown to form
  a complete set of Dirac observables. The action of the asymptotic time
  translation on the observables is defined but it has been calculated
  explicitly only in the case of one-component dust (Vaidya metric).
\end{abstract}


\maketitle


\section{Introduction}

The theory of gravitational collapse and black holes has acquired many facets,
extending from detailed astrophysical models of activity in galactic nuclei to
the sophisticated calculations of the black-hole entropy. Still, some
long-standing central questions persist: Is the outcome of collapse always a
black hole or can in some cases naked singularities be formed? Even if the
cosmic censorship conjecture will sometime be formulated as a theorem, the
final singularities inside black holes will require new physics. Hopes in
curing singularities by a suitable form of quantum gravity are older than the
concept of a black hole itself. Despite new advances towards a general
formulation of quantum gravity, simplified models play still important role,
not least because within such models one may address oneself more easily to a
concrete physical question.

The spherical gravitational collapse of null dust belongs to this type of
prob\-lems.  Since Vaidya's discovery of an exact solution of Einstein's
equations describing a ``radiating spherical star'' \cite{VA}, null dust --
also referred to in literature as ``pure radiation field'' -- has been widely
used as a simple matter source. Its energy-momentum tensor
\begin{eqnarray}\label{a}
T_{\alpha\beta} = \rho k _\alpha k_\beta,
\end{eqnarray}
where $k_\alpha$ is null,
\begin{eqnarray}\label{b}
k_\alpha k^\alpha = 0,
\end{eqnarray}
may be interpreted as an incoherent superposition of waves with random phases
and polarizations moving in a single direction. Null dust exhibits all features
of the geometrical optics limit of Maxwell's theory except for the polarization
properties \cite{MTW}, \cite{BiKu}. In this limit the mass distribution $\rho$
can be identified with the square of the scalar wave amplitude. Alternatively,
the energy-momentum tensor (\ref{a}) can be considered as representing photons
that move along null rays with the flux vector determined by the amplitude and
the null vector $k^\alpha$ \cite{MTW}. The lightlike particles may be massless
scalar particles or neutrinos. The same energy-momentum tensor describes
high-frequency gravitational waves. One can also reinterpret some exact
solutions of Einstein's equations with null dust as spacetimes produced by
zero-rest-mass fields. All these roles of null dust indicate that it is more
closely related to fundamental matter fields than ordinary dust from
phenomenological massive particles. For more details and references on the null
dust and geometrical optics, see Appendix A in \cite{BiKu}; for the list of
various exact solutions with null dust, see \cite{SK}, and Appendix B in
\cite{BiKu}, where some applications are also mentioned.

Vaidya metric is an interesting model in respect of studying the singularities
and their possible removal by quantum phenomena. Indeed, classical solutions
with null dust contain both naked and ``clothed'' singularities. In the next
Section 2 we shall briefly review how spherical null dust solutions can be
used to study the formation of naked singularities, or to describe mass
inflation inside black holes.

The main purpose of the present work is to construct the canonical theory of
spherically symmetric spacetimes with null dusts. The general Lagrangian and
Hamiltonian framework which includes (single-component) null dust as a source
into canonical gravity was developed in Ref.\ \cite{BiKu}.  In Sec.\ 3, we
generalize the Hamiltonian action given in \cite{BiKu} to include both
radially ingoing and outgoing null dusts. We assume that the null dust
components interact only gravitationally, i.e., even in the region where the
two streams interpenetrate, the energy-momentum tensors of both components are
conserved separately. In the same section, the action is reduced by spherical
symmetry. Since in this paper we are interested in the interaction of outgoing
and ingoing null dusts, let us add that recently the similarity solutions
representing interacting null dusts (not necessarily with the same type of
interaction between the components as we have chosen) were studied \cite{Hol};
and the exact static solution for two colliding spherically symmetric null
dust streams in equilibrium was presented \cite{Ge}.

Although the present paper does not address any quantum problem directly, our
main motivation for this generalization originates in some problems of the
quantum theory of gravitational collapse. In Refs.\ \cite{HKie} and \cite{H},
quantum gravitational collapse of a spherically symmetric null-dust thin shell
has been studied (this is a kind of limit of the Vaidya metric). The quantum
mechanics constructed there is unitary and describes a bounce and re-expansion
of the shell in spite of the fact that the corresponding classical solution
contains a horizon and a singularity inside it. The problem mentioned above is:
what is the nature of the quantum spacetime geometry around the quantum shell?
The metric itself is not a gauge invariant quantity in the sense that different
gauges lead to unitarily inequivalent metric operators (see, e.g.,
\cite{HKij}). We have proposed, therefore, to study the quantum theory of a
system consisting of two shells so that the second shell can probe the geometry
around the first one and vice versa, carrying some information about it to the
infinity where it may be deciphered by the asymptotic observes. The
investigations of such a system \cite{HAKO} has revealed a rich set of states
in which the shells cross. Such crossing can be described by a finite-step
canonical transformation in spite of the crossing process being instantaneous.
The reason, of course, is that the shells are infinitesimally thin but carry
finite energy and momenta. The canonical transformation describing a crossing
is rather involved and there is no obvious way of how it can be ``quantized.''
(Usually, we quantize an infinitesimal canonical transformation by making the
function that generates it to an operator.) As yet, we have not managed in a
direct way to find a suitable quantum version (i.e., a unitary operator) of the
canonical transformation for the crossing. One hope, however, is that it could
be found indirectly, if we study some limit of the system of two null-dust {\em
thick} shells crossing each other: there must be a well-defined Hamiltonian for
this process. Such a study is initiated by the present work.

The gravitational part of the constraints can be simplified by Kucha\v{r}'s
canonical transformations introduced originally in the vacuum Schwarzschild
geometrodynamics \cite{KK} and used subsequently for a null-dust thin shell
model in \cite{LOU}. In Kucha\v{r}'s procedure the quasilocal mass plays an
important role. In case of colliding spherical layers of null dust the
quasilocal mass is well identified in Bondi-type coordinates used by Bardeen
\cite{B} to study the effects of back reaction of the Hawking radiation. These
coordinates will be used extensively throughout the paper.  The Kucha\v{r}
transformation is carried out in Sec. 4, based on a careful analysis of the
resulting boundary terms.

In Sec.\ 5, we reduce the action by a gauge choice so that the Hamiltonian for
cross-streaming null dusts can be found. The gauge is shown to be regular, and
the Hamiltonian is calculated. The result, Eq.\ (\ref{17,4}), is
non-local---one must be careful in calculating its variation. Then the
dynamical equations implied by Hamiltonian (\ref{17,4}) are obtained and, in
Sec.\ 6, compared with the Einstein equations in the form that has been given
by Bardeen \cite{B}.

In Sec.\ 7, we analyze the dynamical equations looking for integrals of motion.
Two explicit integrals of motion per point are found. From them a complete set
of Dirac observables is constructed. The transformation between the original
phase space variables and the Dirac observables so constructed is, however, not
known in an explicit form. Similarly, the expression of the Hamiltonian in
terms of the Dirac observables is not known explicitly. Nevertheless, the
transformation properties, as well as the physical meaning of the Dirac
observables are available. And the lack of the explicit algebra and dynamics of
our Dirac observables does not obstruct our project of finding the quantum
version of the shell crossing. We can work with the canonical coordinates of
Sec.\ 5 and with the Hamiltonian (\ref{17,4}). This will be done in the
subsequent work.

It turns out that the explicit transformation between the Dirac observables
and the variables of the reduced theory, as well as the expression of the
Hamiltonian in terms of the Dirac observables, can be found in the special
case of {\em one}-component dust. This is shown in Sec.\ 8. To achieve this
aim, we start from the extended phase space and constraints (before the
reduction by a gauge choice of Sec.\ 5). We use the Vaidya coordinates as our
covariant gauge condition and introduce the corresponding embedding variables.
The coordinates in the extended phase space are transformed to the embedding
variables and the Dirac observables. Then the reduction to the physical phase
space spanned by the Dirac observables is carried out and all desired
transformations are written down explicitly. In this way the action is
transformed into the Kucha\v{r} form \cite{HKij}---cf.\ Eq.\ (\ref{55,2}).
This section can also illustrate how a nontrivial dynamics of Dirac
observables based on the asymptotic symmetries can be constructed. These ideas
are explained thoroughly in Ref.\ \cite{Y}.

The canonical theory of \textit{one}-component spherical null dust, starting
from the general formalism of \cite{BiKu} and Kucha\v{r}'s transformations,
was recently developed by Vaz \textit{et al} \cite{VaWi}. The authors also
discussed the quantization procedure. However, their final form of the action
is not in the Kucha\v{r} form and their theory is not gauge invariant.

\section{Null dust and formation of naked singularities}
As we have already stated in the Introduction, the null dust models have been
used to clarify, in classical context, the formation of naked singularities
during a spherical gravitational collapse. It has been known for about 30
years that during spherical collapse of ordinary dust, the shells of dust may
cross under suitable condition to generate the density singularity outside the
event horizon (see e.g. \cite{Jos} for a review). Since it is possible to
determine the motion of dust through the singularity, this ``shell-crossing
singularity'' is considered as relatively innocent. More serious is the
``shell-focusing singularity'' which may form at the center of a collapsing
spherically symmetric inhomogeneous dust cloud described by a Tolman-Bondi
metric. When collapse to the center proceeds sufficiently slowly, a
``past-null'' singularity which is visible from infinity may arise at the
center \cite{ES}. Such singularity is a ``strong curvature singularity'', as
defined e.g. in \cite{Tip}.

That a shell focusing naked singularity may form in case of imploding null
dust, was first noticed by Hiscock \textit{et al} \cite{HI}, and analyzed in
detail by Kuroda \cite{KUR} and Papapetrou \cite{Pa}. Hollier \cite{HOL}
proved that this singularity is a strong curvature singularity. A number of
more recent papers analyzed this problem within classical theory \cite{SIN}.
To describe the results and to introduce in more detail the following sections,
we first need a few facts about the Vaidya metric.

The energy-momentum tensor (\ref{a}) is the ``primary'' measurable quantity.
Since the lightlike normalization (\ref{b}) is preserved by an arbitrary
scaling of $k_\alpha$,
\begin{eqnarray}\label{c}
\overline{k}^\alpha = \lambda k^\alpha,\;\;\; \lambda > 0,
\end{eqnarray}
we can simultaneously rescale the ``density'' $\rho$,
\begin{eqnarray}\label{d}
\overline{\rho} = \lambda^{-2} \rho,
\end{eqnarray}
so that the form (\ref{a}) of the energy-momentum tensor is preserved. By
taking $\lambda = \rho^{1/2}$ we can eliminate the scalar $\rho$ and
give the energy-momentum tensor (\ref{a}) just in terms of a single null
vector,
\begin{eqnarray}\label{e}
l^\alpha = \rho^ {1/2} k^\alpha ,
\end{eqnarray}
in the form
\begin{eqnarray}\label{f}
T^{\alpha\beta} = l^\alpha l^\beta .
\end{eqnarray}
This choice maximally simplifies the canonical description of null dust. It
will be employed in the following.

The simplest form of the Vaidya metric describing spherical implosion of
\mbox{null} dust is achieved by using the incoming null coordinate $V,$
together with the Schwarzschild curvature coordinate $R$:
\begin{eqnarray}\label{g}
ds^2 = -\left(1-\frac{2M(V)}{R}\right)dV^2 + 2dV dR + R^2 d\Omega ^2.
\end{eqnarray}
Here $M(V)$ -- the ``advanced mass'', measured at past null infinity
${\cal{I}}^-$ -- may be an arbitrary, non-decreasing function. In the
time-reversed case of outgoing spherical cloud of null dust, the simplest
description is provided in terms of the outgoing null coordinate $U$:
\begin{eqnarray}\label{h}
ds^2 = - \left(1- \frac{2M(U)}{R}\right) dU^2 - 2dU dR + R^2 d\Omega ^2,
\end{eqnarray}
where the ``retarded mass'' $M(U)$, measured at future null infinity
${\cal{I}}^+$, may be an arbitrary, non-increasing function. For $M=$
constant, the metrics (\ref{g}) and (\ref{h}) go over into the Schwarzschild
metric in the ingoing, respectively outgoing Eddington-Finkelstein coordinates
\cite{MTW}. The metrics remain simple if, instead of null coordinate $V,$
respectively $U,$ Eddington time-coordinate $T^-_E = V - R$, respectively
$T^+_E = U + R$, is introduced. However, Vaidya's metric becomes more
complicated in the Schwarzschild-type coordinates \cite{VA}.

In case of ingoing null dust the metric (\ref{g}) together with the Einstein
equations imply
\begin{eqnarray}\label{i}
T^-_{\alpha\beta} = \frac{M_{,V}}{4\pi R^2} k_\alpha^-
k_\beta^- = l_\alpha^- l_\beta^- ,
\end{eqnarray}
where $M_{,V} = \partial M/\partial  V \geq 0$,
\begin{eqnarray}\label{j}
k^-_\alpha = - V_{,\alpha}, \;\;\; l^-_\alpha = -
\left(\frac{M_{,V}}{4\pi R^2}\right)^{1/2} V_{,\alpha}\;\; .
\end{eqnarray}
Analogously, for outgoing null dust we get
\begin{eqnarray}\label{k}
T^+_{\alpha\beta} = \frac{M_{,U}}{4\pi R^2}k^+_\alpha k_\beta^+ =
l ^+_\alpha l_\beta^+ ,
\end{eqnarray}
where $M_{,U} = \partial M/\partial U \leq  0$,
\begin{eqnarray}\label{l}
k^+_\alpha = U_{,\alpha} , \;\;\; l_\alpha^+ =
\left(\frac{-M_{,U}}{4\pi R^2}\right)^{1/2} U_{,\alpha}\;\;.
\end{eqnarray}

\begin{figure}[h]
\includegraphics{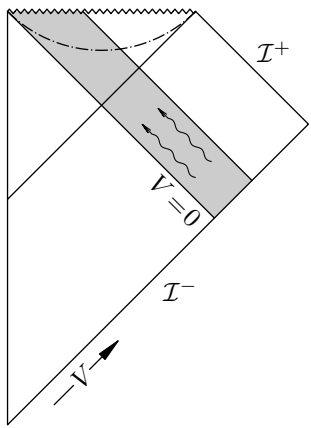}
\caption{The Penrose diagram of a thick spherical layer of null dust
  collapsing from ${\cal{I}}^-$ and forming
  a Schwarzschild black hole. The region containing null dust is shaded, the
  apparent horizon is indicated by dot-and-dashed line, the event horizon by
  full line.}
\end{figure}

\begin{figure}[h]
\includegraphics{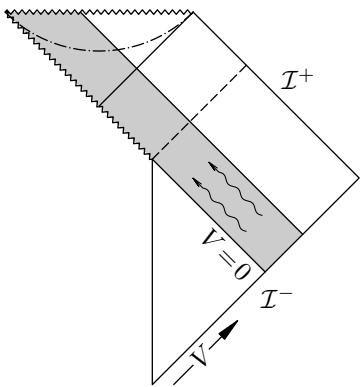}
\caption{The Penrose diagram of a thick spherical layer of null dust
  collapsing from ${\cal{I}}^-$ and forming
  a global naked singularity. The apparent horizon is indicated by
  dot-and-dashed line, the event horizon by full line, and the Cauchy horizon
  by dashed line. (Based on Fig. 3 in \cite{KUR}.)}
\end{figure}

Consider a spherical layer of null dust imploding at $V = 0$ from
${\cal{I}}^-$ (see Figs. 1 and 2). In region $V<0$ spacetime is flat. For
$V\geq 0$ the gravitational field is given by Vaidya's metric (\ref{g}), the
energy-momentum tensor by Eq. (\ref{j}). The mass function $M(V) = 0$ for
$V\leq 0$, $M(V)>0$ is non-decreasing for $V>0$. The scalar invariant quadratic
in the Riemann tensor is equal to $48M^2 (V)/R^6$ -- a naked singularity may
arise at $R=0$, $V\rightarrow 0_+$. If there exists an \textit{outgoing}
radial null geodesic which starts at $R=0$, $V=0_+$ and ends at ${\cal{I}}^+$,
the singularity is globally naked. The analysis of such geodesics is not easy
-- Kruskal-type (``double-null'') coordinates which would enable one a
suitable description of both incoming and outgoing null geodesics can be
introduced explicitly only when $M(V)$ is a linear or exponential function
\cite{LA}. In the linear case, $M(V) = \alpha V, \alpha =$ constant, it can be
proven \cite{KUR} that the singularity will be globally naked if $\alpha\leq
1/16$ and null dust inflow will stop at some $V = V_0$ so that spacetime
becomes Schwarzschild with mass $M(V_0)$. This case is shown in Fig. 2. In the
cases when constant parameter $\alpha > 1/16$, the singularity is hidden
inside the event horizon -- a black is formed as illustrated in Fig. 1.
Therefore, the formation of a naked singularity indeed requires a more gentle
$(\alpha\leq 1/16)$ inflow of matter than the formation of a black hole
$(\alpha > 1/16)$. Recently, naked singularities in higher dimensional Vaidya
spacetimes were also considered \cite{GHD}.

An interesting application of both ingoing and outgoing charged Vaidya metrics
appears in the work of Poisson and Israel \cite{PI} on internal structure of
black holes. An initially static Reissner-Nordstr\"{o}m black hole is
perturbed (non-linearly) by cross-flowing outgoing and infalling null dusts.
The region where the streams intersect lies below the outer horizon.  We refer
a reader in particular to Fig. 3 in \cite{PI} in which -- as in our Fig. 3 in
the following -- there are four regions with different values of the mass
parameter outside cross-flowing streams. Poisson and Israel show that the mass
parameter inflates to classically arbitrarily large values on the Cauchy
horizon inside the black hole. The timelike Reissner-Nordstr\"{o}m
singularity, which is locally naked, is thus probably `converted' into a null
or spacelike singularity. The mass inflation and the instability of the Cauchy
horizon inside black holes have been also studied by other authors \cite{KON}.

The mass inflation represents another classical effect which, in a quantum
context, requires a canonical theory for cross-streaming null dust flows. We
now turn to its formulation.

\section{Canonical formalism for spherically symmetric spacetimes with
  two-component null dust}
We shall start from the geometrodynamical approach to a spherically symmetric
geometry as described in \cite{KK}. Up to very few exceptions (noted in the
following) we adopt the same notation and natural units $G=c=1$.

In geometrodynamics, a spherically symmetric spacetime is
generated by evolving a spherically symmetric three-dimensional
Riemannian space $\Sigma$ with the line element described by two
functions, $\Lambda (r)$ and $R(r)$, of a radial variable $r$:
\begin{equation}\label{1}
d \sigma ^2 = \Lambda ^2 (r) dr^2 + R^2(r) d \Omega ^2.
\end{equation}
Here $d\Omega ^2 = d \theta ^2 + \sin ^2 \theta d \varphi ^2$,
$R(r)$ is the curvature radius of the $2$-sphere $r=$ constant; under
transformations of $r$, $R(r)$ behaves as a scalar, $\Lambda (r)$
as a scalar density. The range of the radial variable $r$ depends
on the problem we wish to discuss. If null dust falls into a
primordial Schwarzschild black hole (which thus existed before an
``irradiation'' by dust), variable $r\in (-\infty, +\infty)$
covers both (``right'' and ``left'') infinities, whereas if an
infalling ``ball'' of null dust creates a Schwarzschild black
hole, the point $r = 0$ is the center of the ball, and $r \in <0,
+\infty)$. As discussed in the Introduction, our main interest will
be in treating the thick shell of null dust falling in or out with
Minkowski space inside. Then again $r \in <0, +\infty).$
Clearly, appropriate boundary conditions must be formulated and
boundary terms analyzed depending on the physical situation
considered.

If a spherically symmetric spacetime is foliated by spherically
symmetric leaves $\Sigma$, labelled by a time parameter $t\in
{\mathbb R}$, the metric functions $\Lambda$ and $R$ depend also on $t$.
The leaves are related by the lapse function $N(r,t)$, and by the
only non-vanishing component of the shift vector $N^r (r,t)$.
Modulo boundary terms, the \textit{vacuum} dynamics of the
spherically symmetric gravitational field follows from the
canonical form of the ADM action, integrated over $\theta$ and
$\varphi$. It reads (cf. \cite{KK}, Eqs. (31)-(33)):
\begin{equation}\label{2}
S^G = \int dt \int dr (P_\Lambda \dot{\Lambda} + P_R \dot{R}
-NH^G - N^r H_r^G),
\end{equation}
where the gravitational super-Hamiltonian
\begin{equation}\label{3}
\begin{split}
H^G = R^{-1}P_RP_\Lambda &+\frac{1}{2} R^{-2} \Lambda P^2_{\Lambda}
+ \Lambda ^{-1} RR'' \\
&- \Lambda ^{-2}RR' \Lambda '
+ \frac{1}{2}\Lambda ^{-1} R'^{2} - \frac{1}{2} \Lambda,
\end{split}
\end{equation}
and supermomentum
\begin{equation}\label{4}
H^G_r = P_RR' - \Lambda P'_\Lambda,
\end{equation}
the dots and primes denoting the derivatives with respect to
variables $t$ and $r$. The momenta, $P_R$ and $P_\Lambda$,
obtained by differentiating the ADM spacetime action with respect
to the velocities $\dot{\Lambda}$ and $\dot{R}$, read
\begin{equation}\label{5}
P_\Lambda = -N^{-1} R (\dot{R} - R' N^r),
\end{equation}
\begin{equation}\label{6}
P_R = - N^{-1} \left[\Lambda(\dot{R} - R'
N^r)+R(\dot{\Lambda}-(\Lambda N^r)') \right].
\end{equation}
Inversely,
\begin{equation}\label{7}
\dot{\Lambda} = -NR^{-2}(RP_{R} - \Lambda P_{\Lambda}) + (\Lambda
N^r)',
\end{equation}
\begin{equation}\label{8}
\dot{R} = -NR^{-1} P_{\Lambda} + R'N^r.
\end{equation}

A general spacetime action for a one-component null dust has the form
\cite{BiKu}
\begin{equation}\label{9}
S^{ND} = \int dt \int d^3x L^{ND},
\end{equation}
where the Lagrangian is given by
\begin{equation}\label{10}
L^{ND} = -\frac{1}{8\pi} \sqrt{|g|} g^{\alpha\beta} l_\alpha l_\beta,
\end{equation}
with
\begin{equation}\label{11}
l_\alpha = W_iZ^i_{,\alpha},
\end{equation}
$i=1,2,3$ is a summation index.
We inserted the factor $(4\pi)^{-1}$ into (\ref{10}) as compared with
the general action introduced in \cite{BiKu}, so as to obtain
simplest expressions after the integration over $\theta$ and
$\varphi$ is made. The symbol $l^\alpha$ denotes the unique vector field
defined by the relation $T_{\mu\nu} = l_{\mu}l_{\nu}$. The fields $Z^i$
represent maps from the spacetime $\mathcal M$ to the material space $\mathcal
Z$ of the dust, each triple $(Z^1,Z^2,Z^3)$ being a name of a particular dust
volume element.

A general coordinate transformation in $\mathcal Z$ can be written as
\begin{equation}
  \bar{Z}^i = \bar{Z}^i(Z^1,Z^2,Z^3), \quad
  \det\left(\frac{\partial\bar{Z}^i}{\partial Z^j}\right) \neq 0 .
\label{2,1}
\end{equation}
For $W_i$, we have $\bar{W}_i = (\partial Z^j/\partial \bar{Z}^i)W_j$, so it
transforms as a covector under the transformation (\ref{2,1}). The Lagrangian
is manifestly invariant. Observe that the transformation group is
Diff$\mathcal Z$, an infinite-dimensional group, but it cannot be considered
as a gauge group. The transformations (\ref{2,1}) are ``rigid'' or ``global''
in the sense that they are independent of spacetime points.

In Ref.\ \cite{BiKu}, the transition to the Hamiltonian action for the null
dust is performed with the result
\begin{equation}\label{3,1}
S^{ND} = \int dt \int d^3x (P_k\dot{Z}^k - N H^{ND} - N^a H^{ND}_a)\ ,
\end{equation}
where
\begin{equation}\label{26}
H^{ND} = \sqrt{g^{ab} H^{ND}_a H^{ND}_b}\ ,
\end{equation}
and
\begin{equation}\label{14}
H^{ND}_a = P_k Z^k_{,a}\ ;
\end{equation}
the variable $P_k$ is the canonically conjugate momentum to $Z^k$ and it
transforms as a covector with respect to (\ref{2,1}); $g^{ab}$ is the inverse
of the metric $g_{ab}$ induced on the leaves $\Sigma$ by the spacetime metric.

During the transition to Eqs.\ (\ref{3,1})--(\ref{14}), important relations
have been obtained and used. Some of them cannot be derived from the reduced
action (\ref{3,1}) any more. Thus, it has to be supplied by the formula for
the vector field $l^\alpha$ which follows from the Hamiltonian equations
(see Ref.\ \cite{BiKu}, Eq.\ (6.29)--(6.31); notice that on the right-hand
side of Eq.\ (6.30) the sign should be ``$+$''):
$$
  l^\alpha = \sqrt{W}n^\alpha - g^{ab}\frac{P_k
  Z^k_{,b}}{\sqrt{|g|W}}Y^\alpha_{,a}\ ,
$$
where $n^\alpha$ is the unit future-oriented normal vector to the foliation
and $Y^\alpha_{,a}$ is determined by the description $Y^\alpha =
Y^\alpha(x^1,x^2,x^3)$ of the foliation in the coordinates $Y^\alpha$ of the
spacetime and $x^k$ of the foliation surfaces. The abbreviations
$$
  W = \sqrt{\frac{g^{ij}P_iP_j}{|g|}}\ ,\quad
  g^{ij} = g^{ab}Z^i_{,a}Z^j_{,b}\ , \quad
  |g| = \mbox{det}g_{ab}\ ,
$$
are used; $g^{ij}$ is the induced metric on $\Sigma$ expressed in the basis
$Z^i_{,a}$.

To reduce action (\ref{3,1}) by spherical symmetry, we choose
$$
   (Z^1,Z^2,Z^3) \equiv (\Phi,\theta,\varphi)\ ,
$$
where $\Phi$ labels the comoving spherical surfaces of the dust while $\theta$
and $\varphi$ mark the radial lines of the dust. From the symmetry, we have
$\dot{\theta} = \dot{\varphi} = \theta' = \varphi' = 0$; we assume that the
spacetime coordinates $\theta$ and $\varphi$ are identical with the dust
coordinates. The conjugate momenta are (cf.\ Eq.\ (6.18) of \cite{BiKu})
$$
  (P_1,P_2,P_3) \equiv (\Pi,0,0)\ .
$$
For the metrics, we have
\begin{equation}\label{33}
\begin{split}
  ds^2 = - \left[ N^2 - \Lambda ^2 (N^r)^2 \right] dt^2 &
  + 2\Lambda^2 N^r dt dr \\
  &\!\!\!+ \Lambda ^2 dr^2 + R^2 d\Omega ^2\ ,
\end{split}
\end{equation}
and
$$
  (N_1,N_2,N_3) = (N^r,0,0)\ .
$$
Then,
$$
  g_{ij} = \mbox{diag}(\Lambda^2\Phi^{\prime -2},R^2,R^2\sin^2\vartheta)\ ,
$$
$$
  W = \frac{|\Pi\Phi'|}{\Lambda\sqrt{|g|}}\ ,
$$
and
\begin{equation}\label{21,1}
  l^\alpha = \frac{\sqrt{|\Pi\Phi'|}}{\Lambda R}\left(n^\alpha -
  \frac{\epsilon}{\Lambda}Y^\alpha_{,r}\right)\ ,
\end{equation}
where $\epsilon = \mbox{sgn}(\Pi\Phi')$ describes the two possibilities of
radially out- and ingoing dust. Thus, we obtain after integration
over the angles:
\begin{equation}\label{4.1}
  S^{ND} = \int dt\,dr \left(\Pi[\dot{\Phi} - (\frac{\epsilon N}{\Lambda} +
  N^r)\Phi']\right)\ .
\end{equation}
As $g_{00} < 0$, we must have $N/\Lambda > N^r$. If $\dot{\Phi}$ is chosen to
be positive, we must have $\epsilon\Phi' > 0$. For ingoing dust, $\Phi' > 0$,
while for the outgoing one, $\Phi' < 0$.  Hence, $\epsilon = +1$ for the in-
and $\epsilon = -1$ for the outgoing dust.

The generalization from one to several dust components that do not directly
interact with each other (indirectly means through the gravitation) is simple:
several terms in the action similar to the one-component action must be added.
This, for two components, leads to the null-dust action of the form:
\begin{equation*}
\begin{split}
  S^{ND}_2 = \int &dt\,dr \;\Bigl[\Pi_+\dot{\Phi}_+ + \Pi_-\dot{\Phi}_-\\
  &-\frac{N}{\Lambda}(|\Pi_+\Phi_+'| + |\Pi_-\Phi_-'|)\\
  &- N^r (\Pi_+\Phi_+' + \Pi_-\Phi_-')\Bigr]\ .
\end{split}
\end{equation*}
The total action is
\begin{equation}\label{7,1}
  S_2 = \int dt dr (P_\Lambda\dot{\Lambda} + P_R\dot{R} + \Pi_+\dot{\Phi}_+ +
  \Pi_-\dot{\Phi}_- - NH - N^rH_r)\ ,
\end{equation}
where
\begin{eqnarray}
  H & = & \frac{P_RP_\Lambda}{R} + \frac{\Lambda P_R^2}{2R^2} +
  \frac{RR''}{\Lambda} - \frac{RR'\Lambda'}{\Lambda^2} + \frac{R^{\prime
  2}}{2\Lambda} - \frac{\Lambda}{2} \nonumber \\
    & + & \frac{|\Pi_+\Phi_+'|}{\Lambda} +
  \frac{|\Pi_-\Phi_-'|}{\Lambda}\ ,
\label{7,2} \\
  H_r & = & P_RR' - \Lambda P'_\Lambda + \Pi_+\Phi_+' + \Pi_-\Phi_-'\ .
\label{7,3}
\end{eqnarray}
Eq.\ (\ref{21,1}) must be written for each dust component defining in our case
the vector fields $l^\alpha_+$ and $l^\alpha_-$, so that
\begin{equation}\label{17.1}
  T^{\alpha\beta} = \frac{1}{4\pi}\left(l^\alpha_+l^\beta_+ +
  l^\alpha_-l^\beta_-\right)\ ,
\end{equation}
the total stress-energy tensor is the sum of the component terms because
the total null-dust action is the sum of two terms of the form (\ref{10}).

\section{Kucha\v{r} transformation}
The gravitational part of constraint (\ref{7,2}) can be simplified by a
canonical transformation that replaces $\Lambda$ by the quasilocal mass $M$ as
it has been first shown by \cite{KK}. One way to define the mass is to use the
coordinates introduced by Bardeen \cite{B}. The metric reads
\begin{equation}\label{29a}
  ds^2 = - Fe^{2\psi}dV^2 + 2e^\psi dV dR + R^2d\Omega^2
\end{equation}
in the {\em advanced} Bardeen coordinates (ABC), and
\begin{equation}\label{29r}
  ds^2 = - Fe^{2\phi}dU^2 - 2e^\phi dU dR + R^2d\Omega^2
\end{equation}
in the {\em retarded} Bardeen coordinates (RBC). Here, we use the abbreviation
\begin{equation}\label{31}
  F = 1 - 2M/R\ ,
\end{equation}
and the pairs of functions $M(V,R), \psi(V,R)$ or $M(U,R), \phi(U,R)$ specify
the geometry. The function $M$ is a scalar field (it is {\em the} quasilocal
mass of the spherically symmetric spacetimes) while $\psi$ and $\phi$ are
potentials determined up to the transformation
$$
  \psi \mapsto \psi + \ln v'\ ,\quad \phi \mapsto \phi + \ln u'\ ,
$$
because the null coordinates $U$ and $V$ are determined up to
$$
  V \mapsto v(V)\ ,\quad U \mapsto u(U)\ ,
$$
$u$ and $v$ being two arbitrary functions of one variable (the prime denoting
the ordinary derivative of a function of one variable).  We shall work with
both ABC and RBC but derive our formulae explicitly only for ABC.

Since we foliate the spacetime by spherically symmetric
spacelike leaves $\Sigma$, labelling the leaves by a time parameter
$t\in {\mathbb R}$, and use a radial label $r$ on the leaves,
we can substitute the foliation
\begin{equation}\label{30}
  V = V(t,r),\,\,\,\,\,\,\,\,  R = R(t,r)
\end{equation}
into the line element (\ref{29a}) and find
\begin{equation}\label{32}
\begin{split}
 ds^2 = \,&  - (F e^{2\psi} \dot{V}^2 - 2 e^\psi \dot{V}\dot{R}) dt^2\\
 & + 2(-F e^{2\psi} \dot{V}V' + e^{\psi}\dot{V}R' + V'\dot{R}) dt dr \\
 & +  (-F e^{2\psi} V'^2 + 2e^{\psi} V'R') dr^2 + R^2 d\Omega^2,
\end{split}
\end{equation}
where the dots and commas again denote $\partial/\partial t$ and
$\partial/\partial r$. Comparing (\ref{32}) with the general ADM form of the
spherically symmetric line element (\ref{33}), we can express function
$\Lambda$, and the lapse and shift functions in the form:
\begin{equation}\label{34}
  \Lambda ^2 = - F e^{2\psi} V'^2 + 2 e^{\psi}V'R',
\end{equation}
\begin{equation}\label{35}
  N = \frac{e^{\psi}(\dot{V}R' - V'\dot{R})}{\Lambda},
\end{equation}
\begin{equation}\label{36}
  N^r = \frac{-F e^{2\psi} \dot{V}V' + e^{\psi}\dot{V}R' +
  e^{\psi}V'\dot{R}}{\Lambda ^2}.
\end{equation}
Returning now back to the expression (\ref{5}) for the momentum
$P_\Lambda$, we may substitute for $\Lambda, N$, and $N^r$ to
obtain
\begin{equation}\label{37}
  P_\Lambda = - \frac{R(F e^{\psi}V' - R')}{\Lambda}\ .
\end{equation}
From the last equation we can express $V'$ regarding (\ref{34}) in the
form
\begin{equation}\label{38}
  -e^{\psi}V' = -\frac{\Lambda P_\Lambda}{FR} + \frac{R'}{F}\ .
\end{equation}
From Eq. (\ref{34}), we can calculate function $F$, Eq.  (\ref{31}), in terms
of the canonical data:
\begin{equation}\label{39}
  F = \left( \frac{R'}{\Lambda} \right) ^2 - \left(
  \frac{P_\Lambda}{R}\right)^2.
\end{equation}
Substituting for $F$ into Eq.\ (\ref{38}), we obtain $e^\psi V'$ also entirely
in terms of the canonical data. Finally, combining Eqs. (\ref{31}) and
(\ref{39}) we can express the mass $M$ as a function of the canonical data,
\begin{equation}\label{40}
  M = \frac{1}{2}R^{-1}P^2_\Lambda - \frac{1}{2}\Lambda ^{-2}RR'^2 +
  \frac{1}{2}R.
\end{equation}
Notice that the equations (\ref{39}) and (\ref{40}) are of exactly the same
form as in the vacuum Schwarzschild geometry, cf.\ \cite{KK}. They can also be
interpreted in a similar way.

Clearly, the corresponding expressions in RBC are obtained by replacing
$$
  e^\psi dV \mapsto -e^\phi dU
$$
everywhere, and by changing the sign of $N$ (keeping $N$ positive).

In vacuum spherically symmetric spacetimes one can find a
canonical transformation to a new set of variables in terms of
which the constraints remarkably simplify \cite{KK}. The
transformation replaces the canonical variables $(\Lambda, P_\Lambda , R,
P_R)$ by new canonical variables $(M, P_M, R, {\cal P}_R)$
where $M$ is given by Eq. (\ref{40}),
\begin{equation}\label{41}
  P_M = R^{-1} F^{-1} \Lambda P_\Lambda,
\end{equation}
\begin{equation}\label{42}
  R = R,
\end{equation}
\begin{equation}\label{43}
\begin{split}
  {\cal P}_R &= P_R - \frac{1}{2}R^{-1} \Lambda P_\Lambda -
  \frac{1}{2}R^{-1}F^{-1}\Lambda P_\Lambda \\
  &-R^{-1}\Lambda ^{-2}F^{-1}\left[ (\Lambda P_\Lambda)' (RR')-(\Lambda
  P_\Lambda)(RR')' \right],
\end{split}
\end{equation}
where $F$ is given in terms of ``old'' canonical variables by
(\ref{39}). This transformation was shown \cite{KK} to be canonical
without employing constraints or dynamical equations. We can make
easily sure that it remains a canonical transformation also on the
extended phase space which includes the null dust canonical
variables $\Phi_+,\Pi_+$ and $\Phi_-,\Pi_-$.

The only point we have to check is that the boundary terms vanish as in the
case of primordial vacuum Schwarzschild black holes analyzed in \cite{KK}. We
consider the following arrangement: An outgoing thick shell of dust starts at
the past singularity and runs through the spacetime to reach ${\mathcal I}^+$;
another thick shell starts at ${\mathcal I}^-$, crosses the first one and
reaches the future singularity (Fig.\ 3). Under both shells, the spacetime is
flat (Schwarzschild mass parameter $M = 0$). In the future of both shells, we
have Schwarzschild spacetime with mass $M = M_3$, in their common past $M =
M_1$, and above both shells, $M = M_2$. The hypersurface $\Sigma$ is assumed
to run from the regular origin $R = 0$ in Minkowski spacetime to the $i^0$ of
the Schwarzschild spacetime above the thick shells, as indicated in Fig.\ 3.

\begin{figure}[h]
\includegraphics{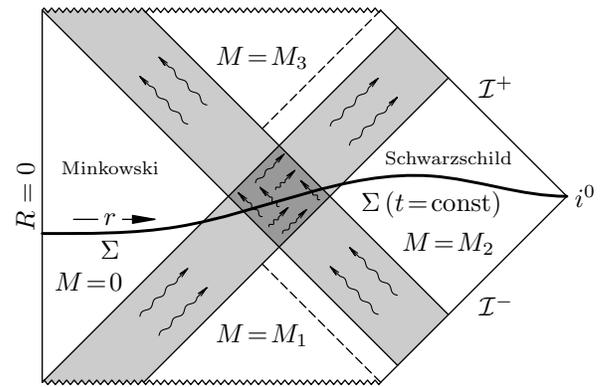}
\caption{The Penrose diagram of two crossing  thick spherical layers (shaded)
  of null dust. One layer is
  collapsing from ${\cal{I}}^-$ into a future Schwarzschild singularity of mass
  $M_3$. The other is expanding from the past singularity of mass $M_1$ into
  ${\cal{I}}^+$. Below the layers the spacetime is flat $(M = 0)$, above the
  layers the Schwarzschild mass is $M_2$. A general Cauchy surface $\Sigma$
  going from the origin $R = 0$ through the crossing region to $i^0$ is
  shown. The past and future event horizons are indicated by dashed lines.}
\end{figure}

The ABC near $R = 0$ of $\Sigma$ must be a transform of the advanced Minkowski
coordinates  $V_{\text{M}},R,\theta,\varphi$:
$$
  V_{\text{M}} = v(V)\ ,
$$
where the function $v$ could in principle be determined by boundary conditions
at infinity and integration of Einstein's equations through the thick shells.
For $V_{\text{M}}$, $R$, $\theta$ and $\varphi$ , the Bardeen potential $\psi$
is zero, hence
$$
  \psi(V) = \ln v'(V)\ .
$$

At ${\mathcal I}^-$, we assume $V$ to be the advanced time. Thus, $V$ is the
Eddington-Finkelstein coordinate above and in the past of the thick shells.
Within the ingoing shell and above the outgoing one, $V$ is the advanced
Vaidya coordinate. We can, therefore, set $\psi = 0$ in a neighbourhood of
${\mathcal I}^-$, and in the part of ${\mathcal I}^+$ lying above the outgoing
shell.

We have thus to pay attention to the behaviour of the canonical variables
$\Lambda$, $R$, $P_\Lambda$ and $P_R$ at the origin and at spatial infinity.
Since we assume that thick shells of radiation do not extend to spatial
infinities, there our fall-off conditions are identical to those considered in
\cite{KK}. Therefore, we assume the canonical variables at infinities to
behave as follows:
\begin{equation}\label{44}
\begin{aligned}
 \Lambda (t,r) &= 1 + M_2 (t)|r|^{-1}+O (|r|^{-(1+\epsilon)}),\\
  R(t,r) &= |r| + O (|r|^{-\epsilon}),  \\
  P_\Lambda (t,r) &= O (|r|^{-\epsilon}),  \\
  P_R(t,r) &= O (|r|^{-(1+\epsilon)}),
\end{aligned}
\end{equation}
where $O(r^{-i})$ means that this term falls off as $r^{-i}$, whereas its j-th
spatial derivatives as $r^{-(i+j)}$; the absolute values appear since in the
case of primordial black holes $r\in {\mathbb R}$. The mass term can depend on
$t$.  The Schwarzschild time $T$ outside the thick shells along a slice
$\Sigma$ is assumed to behave as
\begin{equation}\label{45}
  T(r) = T_\infty (t)+ O(r^{-1}),
\end{equation}
whereas the Lagrange multipliers in the action, the lapse and
shift, satisfy conditions
\begin{equation}\label{46}
  N = N_\pm (t) + O (|r|^{-\epsilon}), \;\;\;\;\;\; N^r =
  O(|r|^{-\epsilon}).
\end{equation}
Since advanced time $V$ is given in terms of canonical coordinates by Eqs.
(\ref{38}), (\ref{39}) with $\psi = 0$, the fall-off conditions (\ref{44})
imply
\begin{equation}\label{47}
  V = |r| + 2M_\infty \ln \left( \frac{r}{2M_\infty} \right) + T_\infty (t) +
  O(r^{-\epsilon}),
\end{equation}
so that
\begin{equation}\label{48}
  V' = 1 + 2M_\infty \frac{1}{|r|}+ O(|r|^{-1-\epsilon}).
\end{equation}
At the origin of Minkowski space the boundary conditions are the same as in
\cite{HKie}. The 4-metric is flat, the 3-metric must be smooth. The
hypersurface $\Sigma$ must meet the origin parallelly to $T =$ constant
hypersurfaces to avoid conical singularities. Therefore, we require at
$r\rightarrow 0$
\begin{equation}\label{49}
  T' (r) = 0, \;\;\;\;\;\; R = r + O (r^2),
\end{equation}
\begin{equation}\label{50}
  V_{\text{M}} = T + R = T_0 + r, \;\;\;\;\; V' = 1.
\end{equation}
The boundary terms, given by the expression (95) in \cite{KK},
\begin{equation}\label{51}
  \mathcal{B} = \frac{1}{2} R \delta R \ln \left|
  \frac{RR' + \Lambda P_\Lambda}{RR' - \Lambda P_\Lambda}\right|,
\end{equation}
evaluated at the boundaries, can also be written in terms of the
advanced time variable $V$ and function $F$ as
\begin{equation}\label{52}
  \mathcal{B} = \frac{1}{2} R\delta R \ln
  \left|\frac{V'e^\psi FR - 2 RR'}{e^\psi V' FR}\right|\ ,
\end{equation}
where $V'e^\psi = V'_{\text{M}}$ at $R =0$ and $V'e^\psi = V'$ at $R
\rightarrow \infty$. Regarding the boundary conditions (\ref{44})--(\ref{50}),
which imply $\delta R \sim O (|r|^{-\epsilon})$ at infinity(ies) and $\delta R
\sim O(r ^2)$ at the origin, we easily see that the boundary terms vanish in
all the situations we are dealing with.

The constraints (\ref{7,2}) and (\ref{7,3}) acquire now much simpler form:
\begin{equation}\label{53}
  \Lambda H = - F^{-1}M'R' - FP_M \mathcal {P}_R + |\Pi_+\Phi'_+| +
  |\Pi_-\Phi'_-| ,
\end{equation}
\begin{equation}\label{54}
H_r = \mathcal {P}_RR' + P_MM' + \Pi_+\Phi'_+ + \Pi_-\Phi'_-\ ,
\end{equation}
where $F$ is again given by (\ref{31}), but $M(r)$ and $R(r)$ are now
understood as new canonical variables. (Notice that in (\ref{53}) we
scaled $H$ by $\Lambda$.) From Eq.\ (\ref{38}) we find the advanced
time coordinate in terms of new canonical variables also in a
simplified form:
\begin{equation}\label{55}
  -e^\psi V' = P_M - F^{-1}R'.
\end{equation}

As we have seen above, we may assume that $\Pi_-\Phi'_- > 0$ for the ingoing
and $\Pi_+\Phi'_+ < 0$ for the outgoing dust. We then combine Eqs.\
(\ref{53}) and (\ref{54}) to obtain:
\begin{equation}\label{newconst}
  H_r \pm \Lambda H = (R' \mp FP_M)({\mathcal P}_R \mp F^{-1}M') + \left\{
  \begin{array}{l}
           2\Pi_-\Phi'_- \\
           2\Pi_+\Phi'_+   \end{array} \right. \ .
\end{equation}
The total action (\ref{7,1})--(\ref{7,3}) can now be written in the form
\begin{eqnarray}
  S &=& \int dt\, dr \left[P_M\dot{M} + {\mathcal P}_R\dot{R} +
  \Pi_+\dot{\Phi}_+ + \Pi_-\dot{\Phi}_- \right. \nonumber \\
      &-& \left. N_+H_+ - N_-H_- - N(\infty)M(\infty)\right]\ ,
\label{newaction}
\end{eqnarray}
where
\begin{eqnarray*}
  H_+ & = & (R' + FP_M)({\mathcal P}_R + F^{-1}M') + 2\Pi_+\Phi'_+ \ ,\\
  H_- & = & (R' - FP_M)({\mathcal P}_R - F^{-1}M') + 2\Pi_-\Phi'_-\ .
\end{eqnarray*}
and
\begin{eqnarray}
  N_+ & = & \frac{1}{2}\left(N^r - \frac{N}{\Lambda}\right)\ ,
\label{13,3} \\
  N_- & = & \frac{1}{2}\left(N^r + \frac{N}{\Lambda}\right)\ .
\label{13,4}
\end{eqnarray}
The action has also been supplied with the ADM boundary term.

\section{Reduction by a gauge choice}
In this section, we shall reduce the action (\ref{newaction}) by choosing
a gauge and by solving the constraints. The equations will simplify strongly,
and this will help us to perform further transformations.

Let us try the gauge defined by the conditions
\begin{equation}\label{15,1}
  r - R = 0\ ,\quad P_M = 0\ .
\end{equation}
We have to check that this is a regular gauge, that is, the gauge functions
defined by the left-hand sides of Eqs.\ (\ref{15,1}) must have vanishing
Poisson brackets with the generators of gauge transformations. From the action
(\ref{newaction}), we obtain
\begin{align}
\dot{R} &= (N_+ - N_-)FP_M + (N_+ + N_-)R'\ ,\label{16,3} \\
\begin{split}
  \dot{P}_M &= [(N_+ + N_-)P_M]' + \left[\frac{R'}{F}(N_+ - N_-)\right]'\\
  &\quad+ \frac{2}{R}(N_+ - N_-)\left(P_M{\mathcal P}_R - \frac{R'M'}{F^2}\right)\ .
\end{split}\label{16,4}
\end{align}
For the Poisson brackets it holds, of course, that $\{R,N_+H_+ + N_-H_-\} =
\dot{R}$ and $\{P_M,N_+H_+ + N_-H_-\} = \dot{P}_M$, where $N_+H_+ + N_-H_-$ is
a generator of gauge transformations for suitable values of $N_+$ and $N_-$.
Regarding Eqs.\ (\ref{16,3}) and (\ref{16,4}) we see that these will be zero
everywhere at the gauge surface only if
$$
  N_+ + N_- = 0\ ,
$$
and
$$
  \left(\frac{N_+ - N_-}{F}\right)' - \frac{2M'}{RF^2}(N_+ - N_-) = 0\ .
$$
These equations are equivalent to
\begin{eqnarray}
  N_+ &=& -KF\exp\left(2\int dR\ \frac{M'}{RF}\right)\ ,
\label{16,1} \\
  N_- &=& KF\exp\left(2\int dR\ \frac{M'}{RF}\right)\ ,
\label{16,2}
\end{eqnarray}
where $K$ is a constant.

The constraints can be solved for ${\mathcal P}_R$ and $M'$:
\begin{eqnarray}
  {\mathcal P}_R &=& -\Pi_+\Phi'_+ - \Pi_-\Phi'_-\ ,
\label{17,1} \\
  M' &=& F(-\Pi_+\Phi'_+ + \Pi_-\Phi'_-)\ .
\label{17,2}
\end{eqnarray}
Eq.\ (\ref{17,2}) shows that $M' \geq 0$ for $F > 0$ while $M' \leq 0$ for $F
< 0$. Moreover, the integrands in Eqs.\ (\ref{16,1}) and (\ref{16,2}) are
always regular and non negative and they converge for our arrangement: at $R =
0$ and near $R = \infty$, there is no dust. Hence, both $N_+$ and $N_-$ will
be non zero at $R = \infty$, and so the generators for which the Poisson
brackets of the gauge functions vanish are not gauge generators. Hence, the
gauge is regular. Eqs.\ (\ref{16,1}) and (\ref{16,2}) determine also the
values of $N_+$ and $N_-$ for our gauge. Eqs.\ (\ref{13,3}) and (\ref{13,4})
then imply
$$
  N^r = 0\ ,\quad N = 2K\Lambda F\exp\left(2\int_0^R dx\ \frac{h(x)}{x}\right)\
  ,
$$
where
\begin{equation}\label{17,6}
  h(R) = -\Pi_+\Phi'_+ + \Pi_-\Phi'_-\ .
\end{equation}
It follows that the gauge (\ref{15,1}) is $R$-orthogonal one: the $t$-lines are
the curves $R =$ constant and the foliation is orthogonal to them. The
foliation is determined by the second equation of (\ref{15,1}), but the time
function that is to be constant along the folios is not. We can choose the
function, which will be called $T$, by imposing the condition on $N$:
$N(\infty) = 1$. This determines the constant $K$ so that the definitive $N$ is
\begin{equation}\label{22,1}
  N(R) = \Lambda F\exp\left(-2\int_R^\infty dx\ \frac{h(x)}{x}\right)\ .
\end{equation}
Eq.\ (\ref{17,1}) together with the Hamiltonian equation for $\dot{M}$ and the
above relations for the lapse and shift yield:
\begin{equation}\label{18,1}
  \dot{M} = \frac{FN}{\Lambda}(\Pi_+\Phi'_+ + \Pi_-\Phi'_-)\ .
\end{equation}

Eq.\ (\ref{17,2}) can be considered as a differential equation for $M$ if the
function $F$ is written out according to Eq.\ (\ref{31}). This equation can be
easily integrated:
\begin{equation}\label{17,3}
  M(R) = \int_0^R dx h(x)\exp\left(-2\int_x^R dy\ \frac{h(y)}{y}\right)\ .
\end{equation}
The Hamiltonian of the reduced theory is the value of the boundary term in
the action (\ref{newaction}) that can be now calculated from the above
equation with the result:
\begin{equation}\label{17,4}
  H_{\mbox{\footnotesize{red}}} = \int_0^\infty dR\,
  h(R)\exp\left(-2\int_R^\infty dx\ \frac{h(x)}{x}\right)\ .
\end{equation}
The reduced action is, therefore, given by
\begin{equation}\label{17,5}
  S_{\mbox{\footnotesize{red}}} = \int_0^\infty dR \left(\Pi_+\dot{\Phi}_+ +
  \Pi_-\dot{\Phi}_- - H_{\mbox{\footnotesize{red}}} \right)\ .
\end{equation}

The Hamiltonian (\ref{17,4}) is non-local.  We can obtain the canonical
equations for the Hamiltonian $H_{\mbox{\footnotesize{red}}}$ as follows.
First, we vary $H_{\mbox{\footnotesize{red}}}$ with respect to $h(R)$:
\begin{equation*}
\begin{split}
  \delta H_{\mbox{\footnotesize{red}}} =
  \int_0^\infty dR\, \delta h(R)
  &\exp\left(-2\int_R^\infty dx\ \frac{h(x)}{x}\right) \\
  &\quad- 2\int_0^\infty dR\, u'(R)v(R)\ ,
\end{split}
\end{equation*}
where we have introduced the abbreviations
$$
  u(R) = \int_0^R dx\, h(x)\exp\left(-2\int_x^\infty
  dy\ \frac{h(y)}{y}\right)\ ,
$$
and
$$
  v(R) = \int_R^\infty dx\ \frac{\delta h(x)}{x}\ .
$$
It is easy to express $u(R)$ in terms of $N(R)$ and $M(R)$ using relations
(\ref{17,3}) and (\ref{22,1}):
$$
  u(R) = \frac{N(R)M(R)}{\Lambda F}\ .
$$
We can see immediately that $u(0) = 0$ and $v(\infty) =0$ so that the last
integral in $\delta H_{\mbox{\footnotesize{red}}}$ is
$$
  -2\int_0^\infty dR\, u'(R)v(R) = -\int_0^\infty dR\
  \frac{2N(R)M(R)}{\Lambda F}\frac{\delta h(R)}{R}\ .
$$
Collecting all terms, we obtain
\begin{equation}\label{deltaH}
  \delta H_{\mbox{\footnotesize{red}}} = \int_0^\infty dR\, \delta
  h(R)\frac{N(R)}{\Lambda}\ .
\end{equation}
The calculation of the canonical equations is now simple if we substitute for
$h(R)$ Eq.\ (\ref{17,6}):
\begin{eqnarray}
  \dot{\Phi}_- - \frac{N}{\Lambda}\Phi'_- &=& 0\ ,
\label{fi-dot} \\
  \dot{\Phi}_+ + \frac{N}{\Lambda}\Phi'_+ &=& 0\ ,
\label{fi+dot} \\
  \dot{\Pi}_- - \left(\frac{N}{\Lambda}\Pi_-\right)' &=& 0\ ,
\label{pi-dot} \\
  \dot{\Pi}_+ + \left(\frac{N}{\Lambda}\Pi_+\right)' &=& 0\ .
\label{pi+dot}
\end{eqnarray}

The canonical equations (\ref{fi-dot})--(\ref{pi+dot}) together with Eqs.\
(\ref{22,1}) and (\ref{17,3}) determine the dynamics of the dust and the
metric.  For example, Eq.\ (\ref{18,1}) can be derived from Eq.\ (\ref{17,3})
by expressing $\dot{M}(R)$ first as a linear functional of $\dot{h}$, similarly
to the calculation of $\delta M$ in terms of $\delta h$, and then by
transforming $\dot{h}$ with the help of Eqs.\ (\ref{fi-dot})--(\ref{pi+dot}).

The spacetime metric in the coordinates $T$, $R$, $\vartheta$ and $\varphi$
has the form, due to our choice of gauge,
\begin{equation}\label{41,1}
  ds^2 = -N^2dT^2 + \Lambda^2dR^2 + R^2d\Omega^2\ .
\end{equation}
The function $\Lambda$ can be obtained from Eq.\ (\ref{34}). Indeed, the gauge
condition $R = r$ and Eq.\ (\ref{36}) with $N^r = 0$ imply
$$
  -Fe^\psi V' + 1 = 0\ ;
$$
this substituted into Eq.\ (\ref{34}) leads to
\begin{equation}\label{41,2}
  \Lambda = \frac{1}{\sqrt{F}}\ ,
\end{equation}
where $M(R)$ is given by Eq.\ (\ref{17,3}). Eqs.\ (\ref{22,1}) and
(\ref{41,2}) determine the function $N$.

\section{Rederivation of Bardeen's equations}
The variation principle (\ref{7,1}) can be checked by deriving Einstein's
equations in Bardeen's coordinates from it. The equations to be derived have
the form \cite{B}:
\begin{eqnarray}
  \frac{\partial \psi}{\partial R} &=& 4\pi R T_{RR}\ ,
\label{31,1} \\
  \frac{\partial M}{\partial R} &=& -4\pi R^2 T^V_V\ ,
\label{31,2} \\
  \frac{\partial M}{\partial V} &=& 4\pi R^2 T^R_V\ .
\label{31,3}
\end{eqnarray}
They are written in the ABC; analogous formulae hold in the RBC.

To derive Eqs.\ (\ref{31,1})--(\ref{31,3}), we have first to express the
stress--energy tensor using Eqs.\ (\ref{17.1}) and (\ref{21,1}). Transforming
the metrics (\ref{29a}) and (\ref{29r}) into the coordinates $T$ and $R$ by
the embedding relations $V = V(T,R)$ and $U = U(T,R)$, and using the
$R$--orthogonality of our gauge, we obtain
\begin{equation}\label{35,1}
  \dot{V} = \frac{N}{e^\psi\sqrt{F}}\ ,\quad \dot{U} =
  \frac{N}{e^\phi\sqrt{F}}\ ,
\end{equation}
and
\begin{equation}\label{35,2}
  V' = \frac{1}{e^\psi F}\ ,\quad U' = -\frac{1}{e^\phi F}\ ,
\end{equation}
where the formulae for the RBC are also written out for later use; $F$ is
to be calculated from Eqs.\ (\ref{31}) and (\ref{17,3}). The
$R$--orthogonality of the gauge implies, in the ABC,
$$
  Y^\alpha_{,R} = (V',1)\ ,\quad n^\alpha =\nu(1,0)\ ,
$$
where $\nu$ is a normalization factor. Eqs.\ (\ref{29a}) and (\ref{35,2}) yield
\begin{equation}\label{31,4}
  Y^\alpha_{,R} = \left(\frac{1}{e^\psi F}, 1\right)\ ,\quad n^\alpha =
  \left(\frac{1}{e^\psi \sqrt{F}}, 0\right)\ .
\end{equation}
Substituting this into Eqs.\ (\ref{17.1}) and (\ref{21,1}), we obtain the
right-hand sides of (\ref{31,1})--(\ref{31,3}):
\begin{eqnarray}
  T_{RR} &=& -\frac{1}{\pi R^2}\Pi_+\Phi'_+\ ,
\label{32,1} \\
   T^V_V &=& \frac{F}{2\pi R^2}\Pi_+\Phi'_+\ ,
\label{32,2} \\
   T^R_V &=& \frac{e^\psi F^2}{4\pi R^2}(\Pi_+\Phi'_+ + \Pi_-\Phi'_-)\ .
\label{32,3}
\end{eqnarray}
On the other hand, we also find the useful formula
$$
  T^-_{\alpha\beta} = \frac{\Pi_-\Phi'_-}{4\pi
  R^2}F^2e^{2\psi}V_{,\alpha}V_{,\beta}\ ,
$$
and, in an analogous way,
$$
  T^+_{\alpha\beta} = -\frac{\Pi_+\Phi'_+}{4\pi
  R^2}F^2e^{2\phi}U_{,\alpha}U_{,\beta}\ .
$$

Next, the derivatives with respect to ABC that occur on the left-hand sides
of Eqs.\ (\ref{31,1})--(\ref{31,3}) are to be expressed in terms of the
primes and dots. Clearly, we have for any function $X(V,R)$:
$$
  \dot{X} = \left(\frac{\partial X}{\partial V}\right)_R\dot{V}\ ,\quad X' =
  \left(\frac{\partial
  X}{\partial V}\right)_RV' + \left(\frac{\partial X}{\partial R}\right)_V\ ,
$$
so that Eqs.\ (\ref{35,1}) and (\ref{35,2}) imply
\begin{equation}\label{32,5}
  \left(\frac{\partial X}{\partial V}\right)_R =
  \frac{e^\psi\sqrt{F}}{N}\dot{X}\ ,\quad
  \left(\frac{\partial X}{\partial R}\right)_V = X' -
  \frac{1}{N\sqrt{F}}\dot{X}\ .
\end{equation}
Analogous formulae for the RBC and $X(U,R)$ read
\begin{equation}\label{32,5r}
  \left(\frac{\partial X}{\partial U}\right)_R =
  \frac{e^\phi\sqrt{F}}{N}\dot{X}\ ,\quad
  \left(\frac{\partial X}{\partial R}\right)_U = X' +
  \frac{1}{N\sqrt{F}}\dot{X}\ .
\end{equation}
Now, Eqs.\ (\ref{17,2}), (\ref{18,1}), (\ref{32,2}), (\ref{32,3}) and
(\ref{32,5}) yield Eqs.\ (\ref{31,2}) and (\ref{31,3}).

To derive Eq.\ (\ref{31,1}), we start from Eqs.\ (\ref{22,1}), (\ref{41,2}),
(\ref{35,1}) and obtain
\begin{equation}\label{32,4}
  e^\psi\dot{V} = \exp\left(-2\int_R^\infty dx\ \frac{h(x)}{x}\right)\ .
\end{equation}
Taking primes of both sides and using Eq.\ (\ref{32,4}) again gives
$$
  e^\psi\dot{V}\psi' +e^\psi\dot{V}' = e^\psi\dot{V}\ \frac{2h(R)}{R}\ .
$$
To calculate $\dot{V}'$, we substitute for $V'$ from Eq.\ (\ref{35,2}):
$$
  \dot{V}' = \left(\frac{1}{e^\psi F}\right)^. = -\frac{\dot{V}}{e^\psi
  F^2}\left(F\frac{\partial \psi}{\partial V} + \frac{\partial F}{\partial
  V}\right)\ .
$$
Of course, from Eq.\ (\ref{31}) we have
$$
  \frac{\partial F}{\partial V} = -\frac{2}{R}\frac{\partial M}{\partial V}\ ,
$$
so that Eq.\ (\ref{32,5}), together with the last three equations, entails
$$
  \frac{\partial \psi}{\partial R} = \frac{2}{R}\left(h - \frac{1}{e^\psi
  F^2}\frac{\partial M}{\partial V}\right)\ .
$$
Then, Eqs.\ (\ref{17,6}), (\ref{32,5}), (\ref{31,2}) and (\ref{31,3}) yield
Eq.\ (\ref{31,1}).

\section{Dirac observables}
The equations of motion imply conservation laws, two for each dust
component. We can use these conserved quantities as Dirac observables.

Consider Eq.\ (\ref{fi-dot}); with the help of Eqs.\ (\ref{41,2}) and
(\ref{35,1}), it can be rewritten in the form
$$
  \left(\frac{\partial}{\partial T} - e^\psi F\dot{V}\frac{\partial}{\partial
  R}\right) \Phi_- = 0\ .
$$
However, Eqs.\ (\ref{41,2}), (\ref{35,1}) and (\ref{35,2}) also imply that
$$
  \left(\frac{\partial}{\partial T} - e^\psi F\dot{V}\frac{\partial}{\partial
  R}\right) V = 0\ .
$$
Hence, $\Phi_-$ is conserved along $V$--\,curves:
\begin{equation}\label{33,1}
  \left(\frac{\partial \Phi_-}{\partial R}\right)_V = 0\ .
\end{equation}
The same result, of course, follows from applying Eq.\ (\ref{32,5}).

The second integral of motion is connected with the conserved current of Eq.\
(4.10) of \cite{BiKu}. It is interesting that it remains conserved even under
the conditions of cross-streaming dusts. It has the form
\begin{equation}\label{33,2}
  \left(\frac{\partial(e^\psi F\Pi_-)}{\partial R}\right)_V = 0\ .
\end{equation}
To prove the equation, we express the left-hand side as follows:
$$
  \left(\frac{\partial(e^\psi F\Pi_-)}{\partial R}\right)_V =
  \left(\frac{\partial (e^\psi F)}{\partial R}\right)_V\Pi_- + e^\psi
  F\left(\frac{\partial\Pi_-}{\partial R}\right)_V\ ,
$$
and calculate the second term by means of Eqs.\ (\ref{32,5}), (\ref{41,2}),
(\ref{35,1}) and (\ref{pi-dot}):
$$
  \left(\frac{\partial\Pi_-}{\partial R}\right)_V = \Pi'_- - \frac{1}{e^\psi
  F\dot{V}}\dot{\Pi}_- = -\frac{(e^\psi F\dot{V})'}{e^\psi F\dot{V}}\Pi_-\ .
$$
We also obtain from Eqs.\ (\ref{32,5}) and (\ref{35,1}) that
$$
  (e^\psi F\dot{V})' = \dot{V}\left(\frac{\partial (e^\psi F)}{\partial
  R}\right)_V\ .
$$
Then, Eq.\ (\ref{33,2}) follows immediately.

Analogous conservation laws, this time along $U$--lines, hold for the outgoing
dust component:
\begin{equation}\label{36,1}
  \left(\frac{\partial \phi_+}{\partial R}\right)_U = 0\ ,\quad
  \left(\frac{\partial(e^\phi F\Pi_+)}{\partial R}\right)_U = 0\ .
\end{equation}

These conservation laws can be utilized for the construction of a nice set of
Dirac observables for the two-component dust. Let us define two functions
$\zeta_-(V)$ and $\mu_-(V)$ by
$$
  \zeta_-(V) = \Phi_-\ ,\quad  \mu_-(V) = e^\psi F\Pi_-\ ,
$$
as well as two functions $\zeta_+(U)$ and $\mu_+(U)$ by
$$
  \zeta_+(U) = \Phi_+\ ,\quad  \mu_+(U) = e^\phi F\Pi_+\ .
$$
The original null-dust canonical variables can then be expressed by means of
these functions as follows
\begin{gather}
\begin{gathered}
  \Phi_-(R) = \zeta_-(V(R))\ ,\\
  \Pi_-(R)  =\mu_-(V(R))V'(R)\ ,
\end{gathered}\label{34,1} \\
\begin{gathered}
  \Phi_+(R) = \zeta_+(U(R))\ ,\\
  \Pi_+(R)  = -\mu_+(U(R))U'(R)\ ,
\end{gathered}\label{34,2}
\end{gather}
where we have substituted $V'$ or $-U'$ for $e^\psi F$ and $e^\phi F$ with the
help of Eq.\ (\ref{35,2}).

Let us finally define the first pair $v(\zeta_-)$ and $u(\zeta_+)$ of the
desired integrals of motion to be the functions inverse to $\zeta_-(V)$ and
$\zeta_+(U)$:
\begin{equation}\label{37,1}
  \zeta_-(v(x)) = x\ ,\quad \zeta_+(u(x)) = x\ .
\end{equation}
The other pair $\mu_-(\zeta_-)$ and $\mu_+(\zeta_+)$ is then defined by
\begin{equation}\label{37,2}
  \mu_-(\zeta_-) = \mu_-(v(\zeta_-))\ ,\quad \mu_+(\zeta_+) =
  \mu_+(u(\zeta_+))\ .
\end{equation}
The two pairs  form a {\em complete set of Dirac observables}.
We assume that the material space $\mathcal Z$ with coordinates $\zeta_-$ and
$\zeta_+$ consists of two components, each with topology ${\mathbb R}^1$:
$$
  \zeta_- \in {\mathbb R}^1\ ,\quad  \zeta_+ \in {\mathbb R}^1\ .
$$
We also assume that the functions $v(\zeta_-)$ and $u(\zeta_+)$ are
non-decreasing.

Relations (\ref{34,1}) and (\ref{34,2}) determine the transformation between
$(\Pi_-,\Phi_-)$ and $(\mu_-,v)$ only implicitly, and so it is for the ``$+$''
variables. The function $V(T,R)$ that occurs in them is, in fact, a
complicated functional of the canonical variables. The reason is that the
transformation between the coordinates $(V,R)$ and $(T,R)$ is {\em solution
  dependent}, and it can, therefore, be written as
$$
  V = V[\Pi_-(T,R),\Phi_-(T,R),\Pi_+(T,R),\Phi_+(T,R);T,R]\ ,
$$
or
$$
  V = V[\mu_-(\zeta_-),v(\zeta_-),\mu_+(\zeta_+),u(\zeta_+);T,R]\ ,
$$
where the functions $\Pi_-(T,R)$, $\Phi_-(T,R)$, $\Pi_+(T,R)$ and
$\Phi_+(T,R)$ in the first formula are considered as the initial datum for a
solution (including the metric) at the Cauchy surface corresponding to a
time value $T$.

Formulae (\ref{34,1}) and (\ref{34,2}) can, nevertheless, serve to find the
transformation properties and physical meaning of the Dirac observables.
Indeed, $\Pi_\pm$ is a density in the material space as well as at the Cauchy
surface $T$ = constant. Divided by $V'$ or $U'$, it becomes a scalar at the
Cauchy surface and a density in the space with coordinates $V$ or $U$.
Thus
$$
  \mu_+(\zeta_+)\dot{u}(\zeta_+)\,d\zeta_+\ ,\quad
  \mu_-(\zeta_-)\dot{v}(\zeta_-)\,d\zeta_-
$$
are scalars with respect to diffeomorphisms in ${\mathcal Z}_\pm$ and with
respect to general transformations of the coordinates $U$ and $V$.

Let us express the total energy $M$ in terms of the Dirac observables. Eqs.\
(\ref{17,2}), (\ref{35,2}), (\ref{34,1}) and (\ref{34,2}) yield
\begin{eqnarray*}
  M &=& \int_0^\infty dR\,F(-\Pi_+\Phi'_+ + \Pi_-\Phi'_-) \\
    &=& \int_0^\infty dR\ \frac{e^{-\phi}\Pi_+\Phi'_+}{U'} +
        \int_0^\infty dR\ \frac{e^{-\psi}\Pi_-\Phi'_-}{V'} \\
    &=& \int_0^\infty dR\,e^{-\phi}\mu_+\ \frac{d\zeta_+}{dR} +
        \int_0^\infty dR\,e^{-\psi}\mu_-\ \frac{d\zeta_-}{dR}\ ,
\end{eqnarray*}
and we obtain:
\begin{equation}\label{48,1}
  M = \int_{{\mathcal Z}_+}d\zeta_+\, e^{-\phi}\mu_+ +
      \int_{{\mathcal Z}_-}d\zeta_-\, e^{-\psi}\mu_-\ .
\end{equation}
With the correcting factors $e^{-\phi}$ or $e^{-\psi}$, the integrands are
just densities in the material space and the energy is a scalar. However, the
explicit dependence of $\phi$ or $\psi$ on the Dirac observables is difficult
to calculate.

The physical meaning of the quantities $\mu_\pm(\zeta_\pm)$ can be made clear
if we consider the arrangement of two crossing thick shells, Fig.\ 3. In a
neighbourhood of ${\mathcal I}^+$ that is intersected only by the outgoing
thick shell, the solution can be written in the form of Vaidya metric that is
given by Eq.\ (\ref{29r}) with $\phi = 0$ and $M = M(U)$. Then, Eq.\
(\ref{17,2}) implies
$$
  \frac{dM}{dU}U' = -F\Pi_+\Phi'_+\ .
$$
Using Eqs.\ (\ref{35,2}) and (\ref{34,2}) with $\phi =0$, we have
$$
  \mu_+ = \frac{dM}{dU}\frac{du}{d\zeta_+} = \frac{dM}{d\zeta_+}\ ,
$$
and, similarly,
$$
  \mu_- = \frac{dM}{d\zeta_-}
$$
near ${\mathcal I}^-$ for the ingoing thick shell. Hence, $\mu_\pm(\zeta_\pm)$
are {\em asymptotic energy densities in the material space}. The corresponding
variables $u(\zeta_+)$ and $v(\zeta_-)$ must then be defined by the retarded
or advanced times $U$ at ${\mathcal I}^+$ and $V$ at ${\mathcal I}^-$;
$u(\zeta_+)$ is {\em the retarded time at which the material sphere $\zeta_+$
  arrives at ${\mathcal I}^+$} and similarly $v(\zeta_-)$ is {\em the advanced
  time at which the material sphere $\zeta_+$ starts at ${\mathcal I}^+$}.

These simple results suggest that we may obtain more explicit formulae for the
special case of one-component dust. This will be shown in the next section.

\section{Vaidya solution}
Let us set $\Pi_+ = 0$ so that there is only the ingoing component of null
dust; from now on, we shall leave out the index ``$-$''. The case with only
the outgoing component can be treated in an analogous way.

Eqs.\ (\ref{31,1}) and (\ref{32,1}) then imply that $\psi = \psi(V)$ and so
$\psi$ can be transformed away by a suitable choice of coordinate $V$.
Moreover, Eqs.\ (\ref{31,2}) and (\ref{32,2}) show that $M = M(V)$. Thus, we
end up with the Vaidya metric.

The canonical theory of the ingoing dust can start with the action
(\ref{newaction}) written out for $\Pi_+ = 0$:
\begin{eqnarray}
  S &=& \int dt\, dr \left[P_M\dot{M} + {\mathcal P}_R\dot{R} +
  \Pi\dot{\Phi} \right. \nonumber \\
      &-& \left. N_+H_+ - N_-H_- - N(\infty)M(\infty)\right]\ ,
\label{51,1}
\end{eqnarray}
where
\begin{eqnarray}
  H_+ & = & (R' + FP_M)({\mathcal P}_R + F^{-1}M')\ ,
\label{51,2} \\
  H_- & = & (R' - FP_M)({\mathcal P}_R - F^{-1}M') + 2\Pi\Phi'\ .
\label{51,3}
\end{eqnarray}
The variables in the action (\ref{51,1}) can be transformed to embeddings and
Dirac observables (Kucha\v{r} decomposition) as follows.

Since Eqs.\ (\ref{34}) and (\ref{55}) together with $\psi =0$ imply
\begin{equation}\label{57}
  FP_M + R' = \frac{\Lambda^2}{V'}\ ,
\end{equation}
the first factor on the right-hand side of Eq.\ (\ref{51,2}) is generally
non-zero.  Hence the factor can be included in the Lagrange multiplier which
we shall call $N^R$:
\begin{equation}\label{58}
  N^R = N_+(R' + FP_M)\ .
\end{equation}
The constraint equation $H_+ = 0$ implies $\mathcal{P}_R +
F^{-1}M' = 0$; substituting this back into the constraint $H_- = 0$, we
get
\begin{equation}\label{59}
  \left(P_M -\frac{R'}{F}\right)M' + \Pi\Phi' = 0\ .
\end{equation}
An equivalent action can thus be written as
\begin{equation}\label{81}
\begin{split}
  S &= \int dt \int _0^\infty dr
  \Bigl[  P_M \dot{M} + \mathcal{P}_R
  \dot{R} + \Pi\dot{\Phi} \\
  &\qquad\qquad-2N_- \bigl((P_M - R'F^{-1})M' + \Pi \Phi'\bigr)\\
  &\qquad\qquad- N^R ({\mathcal P}_R + F^{-1} M')\Bigr] \\
  &- \int dt\, M_\infty
  \dot{T}_\infty\ ,
\end{split}
\end{equation}
where $F$ is determined in terms of canonical variables by Eq. (\ref{31}). The
Regge-Teitelboim boundary term in the action (\ref{51,1}) acquires the form
$M_\infty \dot{T}_\infty$ after the parametrization at infinity, as it is
analyzed in detail in \cite{KK}.  (There would be another term of this type,
corresponding to the left infinity, if null dust is collapsing into a
primordial Schwarzschild black hole.)

The action (\ref{81}) can be simplified by introducing new canonical momenta
conjugate to $M$ and $R$ by
\begin{equation}\label{82}
  \Pi _M = P_M - F^{-1}R',
\end{equation}
and
\begin{equation}\label{83}
  \Pi _R = {\cal P}_R + F^{-1} M'.
\end{equation}
Substituting this into the Liouville form in (\ref{81}), we find
\begin{equation}\label{84}
\begin{split}
  &\int ^{\infty}_{0} dr \left[ P_M \delta M +
  {\cal P}_R \delta R + \Pi \delta \Phi \right]= \\
  &\qquad=\int ^{\infty}_{0} dr \left[ \Pi _M \delta M + \Pi _R
  \delta R +  \Pi \delta \Phi \right]\\
  &\mspace{110mu}+ \int ^{\infty}_{0} dr \left[ F^{-1} (R' \delta M - M' \delta R) \right].
\end{split}
\end{equation}
One can check by a short calculation that
\begin{equation}\label{85}
\begin{split}
  &F^{-1} (R'\delta M - M'\delta R)= \\
  &\qquad= \left(\frac{R \delta R \ln F}{2}\right)'
  - \delta \left(\frac{R'R \ln F}{2} \right),
\end{split}
\end{equation}
so that the transformation defined by Eqs.\ (\ref{82}) and (\ref{83}) is
indeed canonical, since with our boundary conditions the boundary terms
resulting from the first term on the right-hand side of Eq.\ (\ref{85}) drop
out.

Action (\ref{81}) then takes the form
\begin{equation}\label{86}
\begin{split}
  S &= \int dt \int ^\infty _0
  \Bigl[ \Pi _M \dot{M} + \Pi _R \dot{R} + \Pi \dot{\Phi}  \\
  &\qquad\qquad- 2N_- (\Pi _M M' + \Pi \Phi ') - N^R \Pi _R\Bigr] \\
  &- \int dt\, M_\infty \dot{T}_\infty.
\end{split}
\end{equation}

The next step is to transform the variables $M$ and $\Pi_M$ to $V$ and
$\widetilde{\Pi}_V$. Eqs.\ (\ref{55}) with $\psi=0$ and (\ref{82}) imply
\begin{equation}\label{87}
  \Pi _M = -V'\ .
\end{equation}
Let us define $\widetilde{\Pi} _V$ by
\begin{equation}\label{89}
  M(r) = -\int ^r_0 d\widetilde{r} \widetilde{\Pi} _V (\widetilde{r})\ ,
\end{equation}
so that we have
\begin{equation*}
\begin{split}
  &\int ^\infty _0 dr\, \Pi _M \delta M - M_\infty\delta T_\infty =\\
  &\qquad=\int_0^\infty dr\, V'\int_0^r d\widetilde{r}\delta
  \widetilde{\Pi}_V(\widetilde{r}) - M_\infty \delta T_\infty \\
  &\qquad=\lim_{r\rightarrow\infty}\left[V(r)\int_0^r d\widetilde{r}\delta
  \widetilde{\Pi}_V(\widetilde{r}) \right] \\
  &\qquad\qquad\quad- \int ^\infty _0 dr\, V(r)\delta
  \widetilde{\Pi}_V(r) - M_\infty \delta T_\infty\ .
\end{split}
\end{equation*}
Taking into account the boundary condition (\ref{47}) and Eq.\ (\ref{89}), we
have
\begin{equation*}
\begin{split}
  &\int ^\infty _0 dr\, \Pi _M \delta M - M_\infty\delta T_\infty =\\
  &\quad=\int ^\infty _0 dr\, \widetilde{\Pi} _V(r) \delta V(r) - \delta \int ^\infty
  _0 dr\, V(r)\widetilde{\Pi} _V(r)  \\
  &\qquad- \lim_{r\rightarrow\infty}\left[ \left(r +
  2M_\infty\ln\frac{r}{2M_\infty}\right)\delta M(r)\right] \\
  &\qquad- T_\infty\delta  M_\infty - M_\infty\delta T_\infty\ .
\end{split}
\end{equation*}
Since sufficiently large $r$ always lie outside the matter, we can set $M(r) =
M_\infty$ within the limit so that the limit can be written as follows:
$$
  \lim_{r\rightarrow\infty}\delta \left(rM_\infty + 2
  \int_1^{M_\infty}dM\,M\ln\frac{r}{2M}\right)\ .
$$
Discarding all such total differentials in the Liouville form, we obtain
finally
$$
  \int ^\infty _0 dr\, \Pi _M \delta M - M_\infty\delta T_\infty =
  \int_0^\infty dr \widetilde{\Pi}_V(r)\delta V(r)\ .
$$
Hence, the transformation (\ref{87}) and (\ref{89}) leads to the action
\begin{equation}\label{96}
\begin{split}
  S = \int dt \int ^\infty_0 &dr \Bigl[ \widetilde{\Pi} _V \dot{V} + \Pi _R
  \dot{R} + \Pi \dot{\Phi} \\
  &- N^V (\widetilde{\Pi}_V + \Pi\Phi'/V') - N^R \Pi
  _R \Bigr]\ ,
\end{split}
\end{equation}
where $N^V = 2V'N_-$.  The action has been transformed so that the original
ADM coordinates $\Lambda(r)$ and $R(r)$ have been replaced by the embeddings
$V(r)$ and $R(r)$. We observe that the ADM boundary term does not occur
explicitly any more in the action (\ref{96}); this seems to be a general
property of embedding variables (another example is given in Ref.\
\cite{HKie}). It seems that the role of the asymptotic time can be taken over
by the embedding variables. The constraints in the action (\ref{96}) are not
identical with the conjugate momenta $\Pi_V(r)$ and $\Pi_R(r)$ as it should be
for an action after a Kucha\v{r} decomposition \cite{HKij}. The reason is, of
course, that the remaining variables $\Phi(r)$ and $\Pi(r)$, which describe
the physical degrees of freedom, are not Dirac observables and cannot have
zero Poisson brackets with the constraints. In order to make the Kucha\v{r}
decomposition complete, we have to pass to the Dirac observables $\mu(\zeta)$
and $v(\zeta)$.

Eq.\ (\ref{34,1}) that relates $\Pi$, $\Phi$ and $V$ to $\mu$ and $v$ is valid
for an arbitrary embedding $V(r)$ and $R(r)$:
\begin{equation}\label{54,1}
  \Phi(r) = \zeta(V(r))\ ,\quad \Pi(r) = \mu(V(r))V'(r)\ ,
\end{equation}
but the variable $V(r)$ plays now a different role in the formalism: it is
itself a canonical coordinate in the action (\ref{96}). The problem that
the function $V(r)$ in Eq.\ (\ref{34,1}) has not been explicitly known as a
functional of canonical variables (of the reduced theory) has disappeared.

Variations of the functions $\Phi(r)$, $\zeta(v)$ and $V(r)$ are related by
Eq.\ (\ref{54,1}):
$$
  \delta\Phi = \delta\zeta + \frac{d\zeta}{dv}\delta V.
$$
Hence,
$$
  \Pi\dot{\Phi}dr = \mu\dot{\zeta}V'dr + \Pi\frac{d\zeta}{dv}\dot{V}dr
  = \mu\dot{\zeta}dV + \frac{\Pi\Phi'}{V'}\dot{V}dr\ .
$$
The variations of the functions $\zeta(v)$ and $v(\zeta)$ are related by Eq.\
(\ref{37,1}),
$$
  \delta\zeta + \frac{d\zeta}{dv}\delta v = 0\ ,
$$
so that
$$
  \Pi\dot{\Phi}dr + \widetilde{\Pi}_V\dot{V} = -\mu\dot{v}d\zeta +
  \left(\widetilde{\Pi}_V + \frac{\Pi\Phi'}{V'}\right)\dot{V}dr\ .
$$
If we introduce the momentum $\Pi_V$ by
\begin{equation}\label{55,3}
  \Pi_V = \widetilde{\Pi}_V + \frac{\Pi\Phi'}{V'}\ ,
\end{equation}
then the transformation defined by Eqs.\ (\ref{54,1}) and (\ref{55,3}) is
canonical.  Substituting this into the action (\ref{96}), we obtain finally
\begin{equation}\label{55,2}
\begin{split}
  S = \int &dt\biggl[ -\int_{-\infty}^\infty d\zeta\, \mu(\zeta)\dot{v}(\zeta)\\
  &+ \int_0^\infty dr (\Pi_R\dot{R} + \Pi_V \dot{V} - N^R\Pi_R -
  N^V\Pi_V)\biggr]\ .
\end{split}\raisetag{42pt}
\end{equation}
This action is in the Kucha\v{r} form \cite{HKij}.

\begin{acknowledgments}
J.B. thanks Karel Kucha\v{r} for discussions and interest in canonizing Vaidya
solution soon after Ref.\ \cite{BiKu} was finished. Conversations with Joe
Romano and Madhavan Varadarajan were then also helpful. For the revival of the
problem the hospitality of the Institute for Theoretical Physics, University
of Berne, was crucial. A partial support from the grant GA\v{C}R 202/02/0735
of the Czech Republic, as well as from the Swiss National Foundation and
Tomalla Foundation, Zurich are acknowledged. We thank Miroslav Bro\v{z} for the
help with the figures.
\end{acknowledgments}


\end{document}